\documentclass[intlimits,twoside,a4paper]{article}
\usepackage{amsmath,amssymb}
\usepackage{graphicx}
\usepackage[T2A]{fontenc}
\usepackage[cp1251]{inputenc}

\usepackage{color}

\usepackage{cmpj2}

\issue{2015}{18}{2}{23701}
\doinumber{10.5488/CMP.18.23701}

\title[Theoretical studies of the optical and {EPR} spectra  for {VO}$^{2+}$ in $\text{Na}_{3}\text{C}_{6}\text{H}_{5}\text{O}_{7}\cdot2\text{H}_{2}\text{O}$
single crystal]{Theoretical studies of the optical and {EPR} spectra \\ for {VO}$^{2+}$ in $\text{Na}_{3}\text{C}_{6}\text{H}_{5}\text{O}_{7}\cdot2\text{H}_{2}\text{O}$
single crystal}

\author[C.-Y. Li, G.-X. Wang, X.-M. Zheng]{C.-Y. Li\refaddr{ad1}\thanks{E-mail: cyli1962@126.com}\,, G.-X. Wang\refaddr{ad2}, X.-M. Zheng\refaddr{ad1}}

\addresses{
\addr{ad1}School of Physics and Electronic Information,
Shangrao Normal University, Shangrao Jiangxi 334000, P.R. China
\addr{ad2}  Jiangxi Radio \& TV University, Nanchang
330046, P.R. China
}

\date{Received October 9, 2014, in final form November 28, 2014}
\authorcopyright{Ch.-Y. Li, G.-X. Wang, X.-M. Zheng, 2015}

\begin{document}

\maketitle

\begin{abstract}
On the basis of the perturbation formulas for a $d^{1}$
configuration ion in a tetragonal crystal field, the three optical
absorption bands and electron paramagnetic resonance ({EPR})
parameters ($g$ factors $g_{i}$ and hyperfine structure
constants $A_{i}$ for $i =  \|$ and $\perp$,
respectively) of {VO}$^{2+}$ in
$\text{Na}_{3}\text{C}_{6}\text{H}_{5}\text{O}_{7}\cdot2\text{H}_{2}\text{O}$
single crystals were studied using the
perturbation theory method. By simulating the calculated optical and
{EPR} spectra to the observed values, local structure
parameters and negative signs of the hyperfine structure constants
$A_{i}$ of the octahedral $(\text{VO}_{6})^{8-}$ cluster
in trisodium citrate dehydrate single crystal can be obtained.

\pacs 76.30.Fc, 75.10.Dg, 71.70.ch

\end{abstract}

\section{Introduction}

Electron paramagnetic resonance ({EPR}) has long been
considered as one of the most useful tools for the experimental
study of chemical bonding, structural information of the electronic
and spatial configuration of paramagnetic centers \cite{1,2,3}. The
optical absorption method provides the crystal field parameters and the
energy level structure of the metal ion \cite{4,5}. Thus,
{EPR} and optical spectrum technique are two powerful tools
for studying point symmetry and dynamic properties of the
transition-metal ions in the host crystals. The vanadyl ion (i.e.,
{VO}$^{2+}$), having $3{d}^{1}$ configuration,  is the most
stable cation among a few molecular paramagnetic transition metal
ions that is used extensively as an impurity probe for probing the
site symmetry of the central ion and the bonding nature with the
ligands {EPR} studies \cite{1,2,6,7,8}. Moreover, due to the
strong V=O bonding in {VO}$^{2+}$ ion, most of the
{VO}$^{2+}$ complexes in the crystals posses C$_\textrm{4v}$ symmetry
with both $g$ and $A$ values found to be axially symmetric
\cite{6,7}. For example, Karabulut et al. measured the optical
spectra and {EPR} parameters for $\text{V}^{4+}$ center in
trisodium citrate dehydrate
($\text{Na}_{3}\text{C}_{6}\text{H}_{5}\text{O}_{7}\cdot2\text{H}_{2}\text{O}$;
{TSCD} hereafter) single crystal \cite{8}. As is known, for
$3{d}^{1}$ ({VO}$^{2+}$ or $\text{V}^{4+}$) ion in crystals, an
octahedral complex with a tetragonal compression would give $g_{\|}$
$<$ $g_{\bot}$ $<$ $g_{s}$ and $|A_{\|}|$ $>$ $|A_{\bot}|$
\cite{9,10}, where $g_{s}$ is the free-spin $g$ value of 2.0023. The
observed {EPR} parameters (see table~\ref{tab1}) of
{TSCD}$:${VO}$^{2+}$ agree with the relation. That is to
say, the studied $(\text{VO}_{6})^{8-}$ cluster in {TSCD}
crystal is in tetragonally distorted compressed octahedral. Based
on the crystal-field theory, theoretical calculations of the optical
absorption and {EPR} spectra of
{TSCD}$:${VO}$^{2+}$ were performed using the complete
diagonalization energy matrix method ($\text{CDM}$) by Kalfaoglu
and Karabulut, and the calculated results agree with the
experimental data (see table~\ref{tab1},  Cal$^{a}$.) \cite{11}. However, the
previous treatments fail to connect the optical and {EPR}
spectra with the local structures of the system studied and local
lattice distortion around the Jahn-Teller ion $\text{V}^{4+}$ center
was not taken into account. Consequently, the local structures
information of the $(\text{VO}_{6})^{8-}$ cluster in {TSCD}
crystal was not obtained. Since the analysis of the optical and
{EPR} spectra can provide useful information about electronic
states and local structures of the octahedral $(\text{VO}_{6})^{8-}$
cluster in {TSCD} single crystal, which would be helpful in
understanding the properties of this crystal, further studies on the
above experimental results are of specific scientific and practical
significance. In this work, the optical and {EPR} spectra of
{VO}$^{2+}$ doped {TSCD} crystal are a reasonable
explanation for using the perturbation theory method ($\text{PTM}$) for
a $3{d}^{1}$ ion in tetragonally compressed octahedra. In the
calculated formulas, the tetragonal field parameters ${D_s}$ and
${D_t}$ are estimated from the superposition model which makes it possible
to correlate the crystal-field parameters and hence the EPR
parameters with the tetragonal distortion $\Delta R$ ($= R_{\perp}-R_{\parallel}$) of $(\text{VO}_{6})^{8-}$ cluster
in {TSCD} crystal. Based on the above studies, some useful
information of defect structures for the tetragonal $\text{V}^{4+}$
centers in {TSCD} crystal can be obtained from the EPR and
optical spectra analysis. The negative signs of hyperfine structure
constants $A_{\|}$ and $A_{\bot}$ are also suggested from the
calculations.

\section{Calculation}

For a $3{d}^{1}$ ion in tetragonally compressed octahedra, its
higher orbital doublet $^{2}E_{g}$ of the original cubic case would
split into two orbital singlets $^{2}B_{1g}$($|d_{x^{2}-y^{2}}\rangle$)
and $^{2}A_{1g}$($|d_{z^{2}}\rangle$), while the original lower orbital
triplet $^{2}T_{2g}$ would be separated into an orbital doublet
$^{2}E_{1g}$ ($|d_{xz}\rangle$ and $|d_{yz}\rangle$) and a singlet $^{2}B_{2g}$
($|d_{xy}\rangle$), with the latter lying lowest \cite{12,13}. Therefore,
the three optical absorption bands can be given as:
\begin{eqnarray}
E_{1}&=& E(^{2}B_{2g})\rightarrow E(^{2}B_{1g}) = 10D_q,\nonumber\\
E_{2}&=& E(^{2}B_{2g})\rightarrow E(^{2}E_{1g}) = -3D_s+5D_t,\nonumber\\
E_{3}&=& E(^{2}B_{2g})\rightarrow E(^{2}A_{1g}) = 10D_q-4D_s-5D_t.
\end{eqnarray}

The cubic and tetragonal field parameters ({i.e.}, $D_q$,
$D_s$ and $D_t$) can be determined from the superposition model
\cite{13} and the geometrical relationship of the impurity
$\text{V}^{4+}$ center:
\begin{eqnarray}
D_q &=& \frac{3}{4}\,\overline{A}_{4}(R),\nonumber\\
D_s &=&
\frac{4}{7}\,\overline{A}_{2}(R)\left[\left(\frac{\overline{R}}{R_{\parallel}}\right)^{t_{2}}
-\left(\frac{\overline{R}}{R_{\perp}}\right)^{t_{2}}\right],\nonumber\\
D_t &=&
\frac{16}{21}\,\overline{A}_{4}(R)\left[\left(\frac{\overline{R}}{R_{\parallel}}\right)^{t_{4}}
-\left(\frac{\overline{R}}{R_{\perp}}\right)^{t_{4}}\right].
\end{eqnarray}

Here, $t_{2}\approx 3$ and $t_{4}\approx 5$ are the power-law
exponents in view of the ionic nature of the bonds \cite{14,15,16}.
$\overline{R}$ $=$ $(R_{\parallel}+2R_{\perp})/3$ is the average
impurity-ligand ($\text{V-O}$) distance, where $R_{\parallel}$ is the
($\text{V-O}$) distance along the C$_{4}$ axis, and $R_{\perp}$ is
the bonding length between $\text{V}^{4+}$ and the original planar
oxygen ions. $\overline{A}_{2}(R)$ and $\overline{A}_{4}(R)$ are the
intrinsic parameters with the reference distance $\emph{R}$, where
$\emph{R}$ ($\approx$ $R_{\perp}$ $\approx$ 1.985~{\AA} \cite{17})
is taken for the {VO}$^{2+}$ in cubic field. For $3d^{n}$ ions
in octahedra, the ratio
$\overline{A}_{2}(R)$$/$$\overline{A}_{4}(R)$ is in the range of 9--12 in many crystals \cite{4,12,13,18,19,20}. Here, we take
the mean value, {i.e.}
$\overline{A}_{2}(R)$$/$$\overline{A}_{4}(R)$ $\approx$ 10.5.

Within the high order perturbation theory, the third-order
perturbation formulas of {EPR} parameters ($g$ factors
$g_{\parallel}$, $g_{\perp}$ and the hyperfine structure constants
$A_{\|}$, $A_{\perp}$) for $3d^{1}$ ions in tetragonal symmetry with
the ground state $^{2}B_{2g}$ ($|d_{xy}\rangle$) can be derived as \cite{9,
20}:
\begin{eqnarray}
g_{\parallel}&=&g_{s}-8{k}\frac{\zeta}{E_{2}}-({k}+g_{s})\frac{\zeta^{2}}{E_{2}^{2}}
+4{k}\frac{\zeta^{2}}{E_{1}E_{2}}\,,\nonumber\\
g_{\perp}&=&g_{s}-2{k}\frac{\zeta}{E_{1}}+({k}-g_{s})\frac{\zeta^{2}}{E_{1}^{2}}
-2g_{s}\frac{\zeta^{2}}{E^{2}_{2}}\,,\nonumber\\
A_{\parallel}&=&{P}\left[\left(-\kappa-\frac{4}{7}\right)+\left(g_{\parallel}-g_{s}\right)
+\frac{3}{7}\left(g_{\perp}-g_{s}\right)\right]\,,\nonumber\\
A_{\perp}&=&{P}\left[\left(-\kappa+\frac{2}{7}\right)+\frac{11}{14}\left(g_{\perp}-g_{s}\right)\right]\,.
\end{eqnarray}

In the above formulas, $k$ is the orbital reduction
factor. $\zeta$ and $P$ are, respectively, the spin-orbit coupling
coefficient and the dipolar hyperfine structure parameter for the
center $3d^{1}$ ion (i.e., $\text{V}^{4+}$) in crystals. $\kappa$ is
the isotropic core polarization constant. Generally, the value of
$\kappa$ within the range 0.6--1.0 for {VO}$^{2+}$ in
various crystal \cite{4, 10, 19,20,21,22}, here, we take
$\kappa \approx 0.80$, which is comparable with that $\kappa$ ($\approx 0.72$) obtained by the previous work and can be regarded as
reasonable. Thus, the $g$ factors, especially the anisotropy $\Delta
g$  ($=g_{\parallel}-g_{\perp}$) is connected with the tetragonal
field parameters and hence with the local structure (i.e.,
the relative tetragonal elongation $R_{\parallel}-R_{\perp}$)
of the studied systems. Due to the covalence reduction effect
for $3d^{n}$ ions in crystals, we have \cite{19,20,21,22}
\begin{eqnarray}
\zeta = N^{2}\zeta_{0} \qquad \text{and} \qquad P=N^{2}P_{0}\,.
\end{eqnarray}

Here, $N$ ($\approx {k}$) is the covalency reduction
factor characteristic of the covalency effect of the
systems studied. Thus, the spin-orbit coupling coefficient $\zeta$ and the
dipolar hyperfine structure parameter $P$ can be acquired for the
studied systems by using the free-ion data $\zeta_{0}$ ($\approx 248$~cm$^{-1}$ \cite{23}) and $P_{0}$ ($\approx 172 \times 10 ^{-4}$~cm$^{-1}$ \cite{24}) for $\text{V}^{4+}$ ion. So, in the above
formulas, only the parameters $N$, $\overline{A}_{4}(R)$ and
$R_{\parallel}$ are unknown. By fitting the calculated optical and
{EPR} spectra to the experimental values, one can obtain
\begin{equation}
{N}\approx 0.855, \qquad \overline{A}_{4}(R)\approx 1277~\text{cm}^{-1}
\qquad \text{and} \qquad R_{\parallel}\approx 1.733~\text{\AA}.
\end{equation}

The calculated results are compared with the experimental
values in table~\ref{tab1}. Interestingly, the obtained $\text{V-O}$ distance
along the C$_{4}$ axis {V--O} bond lengths ($R_{\parallel}$
$\approx$ 1.733~{\AA}) of the studied $(\text{VO}_{6})^{8-}$ cluster
in {TSCD} crystal, which is very close to that
$R_{\parallel}$ ($\approx 1.739$~{\AA} \cite{4} and 1.60~{\AA}
\cite{20}) for {VO}$^{2+}$ doped in calcium aluminoborate
glasses ($\text{CaAB}$) and
$\text{Cd}\text{Na}\text{P}\text{O}_{4}\cdot6\text{H}_{2}\text{O}$
crystal with similar tetragonally compressed $(\text{VO}_{6})^{8-}$
cluster can be regarded as reasonable.

\begin{table}[!t]
\caption{ The calculated and experimental optical spectra (in
$\text{cm}^{-1}$) and {EPR} parameters $g$ factors and the
hyperfine structure constants (in $10^{-4}~\text{cm}^{-1}$) for
{TSCD}$:${VO}$^{2+}$ cryatal. \label{tab1}}
\vspace{2ex}
\begin{center}
  \begin{tabular}{|c|c|c|c|c|c|c|c|}
   \hline\hline
            &$g_{\parallel}$   &$g_{\perp}$ &$A_{\parallel}$  &$A_{\perp}$  &$^{2}B_{2g}\rightarrow {^{2}E_{1g}}$   &$^{2}B_{2g}\rightarrow {^{2}B_{1g}}$   &$^{2}B_{2g}\rightarrow {^{2}\!A_{1g}}$  \\  \hline \hline
  Cal$^{a}$.       &$1.932$           & $1.992$    &$-182.4$         &$-65.3$       &12168                                   &16873                                           &23957                                    \\     \cline{1-8}
  Cal$^{b}$.       &$1.939$           & $1.988$    &$-180.9$         &$-66.1$       &12410                                   &16893                                           &24625                                    \\     \cline{1-8}
  Expt.[8]         &$1.938$           & $1.998$    &$\phantom{-}183.7$          &$\phantom{-}64.4$        &12195                                   &16892                                           &24631
\\     \cline{1-8}
  \hline\hline
  \end{tabular}
\begin{minipage}{0.9\textwidth}
\vspace{2mm}
{\small $^a$ Calculated by Kalfaoglu and Karabulut using the
complete diagonalization energy matrix method (CDM) in reference \cite{11}.}\\
{\small $^{b}$ Calculated in this work based on the perturbation theory
method ({PTM}).}
\end{minipage}
\end{center}
\end{table}


\section{Discussion}

1) From table~\ref{tab1}, one can find that the calculated (Cal$^{b}$.)
optical and {EPR} spectra are in reasonable agreement with
the observed values. Thus,  the optical and {EPR} spectra of
the system studied are quantitatively interpreted and the local defect
structure of the octahedral $(\text{VO}_{6})^{8-}$ cluster in
{TSCD} crystal is established. The results $R_{\perp}$ $>$
$R_{\parallel}$ indicate that the studied $(\text{VO}_{6})^{8-}$
cluster exists in a tetragonal distortion octahedral site compressed
along the C$_{4}$-axis, which is consistent with the experimental
{EPR} results ({i.e.}, $g_{\parallel}<g_{\perp} < g_{s}$ and $|A_{\|}| > |A_{\bot}|$) under the ground state
$B_{2g}$.

2) The validity of the adopted covalency factor $N$ can be
further illustrated by the relationship $N^{2} \approx  1 -
h(L)k(M)$ \cite{25}, where the parameters $h(L)$ ($\approx 1$) is
the characteristic of the ligands $L$ ($=\text{O}^{2-}$), and
$k(M)$ is the characteristic of the central metal ion \cite{26}.
From the values $k(V^{2+}) \approx 0.1$ \cite{26} and $k(V^{3+})
\approx 0.2$ \cite{26}, one can here reasonably obtain $k(V^{4+})
\approx 0.3$  by extrapolation. Thus, we have $N$ of about
0.84, which is close to the adopted ones ($N \approx 0.855$)
and can be regarded as reasonable.

3) The above calculations suggested that the hyperfine
structure constants $A_{\|}$ and $A_{\bot}$ of {VO}$^{2+}$ in
{TSCD} crystal are negative (see table~\ref{tab1}), but the observed
values given in reference~\cite{8} are positive. It should be pointed out that
the signs of constants $A_{\|}$ and $A_{\bot}$ cannot be determined
solely from {EPR} measurement \cite{27}. Thus, the
experimental values of constants $A_{\|}$ and $A_{\bot}$ are
actually absolute values \cite{6,7,8}. In this paper, we found,
that the signs of $A_{\|}$ and $A_{\bot}$ for {VO}$^{2+}$ in
{TSCD} crystal should be negative, and this is consistent
with a number of theoretical investigations \cite{4, 9, 19,20,21,22}
and the previous calculation \cite{11}. Therefore, the above
calculated $A_{\|}$ and $A_{\bot}$ of the hyperfine structure
constants for {TSCD}$:${VO}$^{2+}$
are reasonable in signs and in magnitude.

\section{Summary}

The optical and {EPR} spectra of the tetragonal
$\text{V}^{4+}$ center in {TSCD} single crystal are
theoretically investigated based on $\text{PTM}$ and related equations,
the calculated results are in good agreement with the observed
values. Large tetragonal compressed distortion $\Delta R$ ($=
0.252$~{\AA}) of $(\text{VO}_{6})^{8-}$ cluster in {TSCD}
single crystal are reasonably established. Obviously, the calculated
method is also effective for other similar
systems.


\section*{Acknowledgements}

This work was financially supported by Chinese Natural Science
Foundation (grant~$11365017$).



\clearpage

\ukrainianpart

\title{Теоретичні дослiдження оптичних спектрів та спектрiв ЕПР для  {VO}$^{2+}$ у монокристалі $\text{Na}_{3}\text{C}_{6}\text{H}_{5}\text{O}_{7}\cdot2\text{H}_{2}\text{O}$}

\author{Чао-Їнг Лі\refaddr{ad1}, Гуо-Кеу Ванг\refaddr{ad2}, Кеу-Меі Женг \refaddr{ad1}}

\addresses{
\addr{ad1}Школа фізики та електронної інформації, педагогічний університет Шанграо, \\ Шанграо Цзянсі  334000, КНР
\addr{ad2}  Університет радіо і телебачення Цзянсі, Наньчан
330046, КНР
}

\makeukrtitle
\begin{abstract}
 Базуючись на пертурбацiйних формулах для iону конфiгурацiї $d^1$ у
 тетрагональному кристалічному полi, було дослiджено три смуги оптичного
 поглинання та параметри електронного парамагнiтного резонансу (ЕПР)
 ($g$-фактори $g_i$ та константи надтонкої структури $A_i$ для $i = \parallel$
 i $\perp$, вiдповiдно) для VO$^{2+}$ у монокристалах
 Na$_3$C$_6$H$_5$O$_7\cdot2$H$_2$O, використовуючи метод теорiї збурень.
 Допасовуючи обчисленi оптичнi спектри та спектри ЕПР до спостережуваних
 значень, можна отримати локальнi структурнi параметри та вiд’ємнi знаки
 констант надтонкої структури $A_i$ октаедричного кластера (VO$_6$)$^{8-}$ в
 монокристалi тринатрійцитрат дигідрату.

\keywords ЕПР, оптичнi спектри, ванадил, тринатрійцитрат дигідрат

\end{abstract}
\end{document}